\documentclass[12pt,preprint]{aastex}

\shorttitle{Heavy Element abundances in 47 Tuc}
\shortauthors{Wylie, Cottrell, Sneden}

\begin{document}

%% LaTeX will automatically break titles if they run longer than
%% one line. However, you may use \\ to force a line break if
%% you desire.

\title{Heavy Element Abundances in AGB stars in 47 Tucanae}

%% Use \author, \affil, and the \and command to format
%% author and affiliation information.
%% Note that \email has replaced the old \authoremail command
%% from AASTeX v4.0. You can use \emaigiml to mark an email address
%% anywhere in the paper, not just in the front matter.
%% As in the title, use \\ to force line breaks.

\author{E. C. Wylie\altaffilmark{1} and P. L. Cottrell}
\affil{Department of Physics and Astronomy, University of Canterbury, \\
    Christchurch, New Zealand}
\altaffiltext{1}{e.wylie@phys.canterbury.ac.nz}

\author{C. A. Sneden\altaffilmark{2}}
\affil{University of Texas, Austin, Texas, USA}
\altaffiltext{2}{Visiting Erskine Fellow, University of Canterbury}

\and

\author{J. C. Lattanzio}
\affil{Centre for Stellar and Planetary Astrophysics, School of Mathematical Sciences, Monash University, 3800, Australia}

%% Notice that each of these authors has alternate affiliations, which
%% are identified by the \altaffilmark after each name.  Specify alternate
%% affiliation information with \altaffiltext, with one command per each

%% Mark off your abstract in the ``abstract'' environment. In the manuscript
%% style, abstract will output a Received/Accepted line after the
%% title and affiliation information. No date will appear since the author
%% does not have this information. The dates will be filled in by the
%% editorial office after submission.

\begin{abstract}
This research forms part of an investigation into heavy element abundances in Asymptotic Giant Branch (AGB) stars in various stellar environments. Seven giant stars in the southern globular cluster 47 Tuc have been observed using the Anglo-Australian Telescope.  Abundances for five s- and r- process elements have been determined: the light s-process elements, Y and Zr; the heavy s-process elements, La and Nd; and the r-process element Eu. Mean enhancements in the light s-process, [ls/Fe], of $\sim$+0.6 dex and heavy s-process elements, [hs/Fe], of $\sim$+0.3 dex were determined for all the giant stars.  There was no statistically significant difference between the abundances determined for the Red Giant Branch (RGB) and AGB stars in this study.  The results for the RGB stars differ from those obtained by a number of previous studies.  However, because of the similar abundance results obtained for the AGB and RGB stars in this study we believe this provides evidence for previous enrichment of the material from which these stars formed.
\end{abstract}

%% Keywords should appear after the \end{abstract} command. The uncommented
%% example has been keyed in ApJ style. See the instructions to authors
%% for the journal to which you are submitting your paper to determine
%% what keyword punctuation is appropriate.

%% Authors who wish to have the most important objects in their paper
%% linked in the electronic edition to a data center may do so in the
%% subject header.  Objects should be in the appropriate "individual"
%% headers (e.g. quasars: individual, stars: individual, etc.) with the
%% additional provision that the total number of headers, including each
%% individual object, not exceed six.  The \objectname{} macro, and its
%% alias \object{}, is used to mark each object.  The macro takes the object
%% name as its primary argument.  This name will appear in the paper
%% and serve as the link's anchor in the electronic edition if the name
%% is recognized by the data centers.  The macro also takes an optional
%% argument in parentheses in cases where the data center identification
%% differs from what is to be printed in the paper.

\keywords{s-process, AGB stars, globular clusters: general ---
globular clusters: individual(\objectname{NGC 104},
\object{47 Tucanae})}

%% From the front matter, we move on to the body of the paper.
%% In the first two sections, notice the use of the natbib \citep
%% and \citet commands to identify citations.  The citations are
%% tied to the reference list via symbolic KEYs. The KEY corresponds
%% to the KEY in the \bibitem in the reference list below. We have
%% chosen the first three characters of the first author's name plus
%% the last two numeral of the year of publication as our KEY for
%% each reference.

\section{Introduction\label{intro}}

The AGB phase is a very short stage of
stellar evolution.  However, the s-process nucleosynthesis that occurs during
this phase is of great consequence in the understanding of element
formation and recycling in the cosmos.  Nucleosynthesis is also an important tracer of temperature and
mixing in stars and, consequently, understanding the nucleosynthesis that occurs during this phase is a vital tool in
astrophysics \citep{1999ARA&A..37..239B}.

For most of the AGB evolution, the He-burning shell (hereafter just the He shell) remains inactive.  However, periodically, the shell undergoes a thermal runaway and generates enormous quantities of energy for very short periods of time, resulting in a thermal pulse, or shell flash.  AGB stars experience many recurring thermal pulses during their
lifetime and it is the nucleosynthesis, heavy element formation and mixing (Third Dredge Up, TDU) occurring
throughout this phase that has implications for this research.

Globular clusters provide an excellent environment for studying stellar evolution.  It is assumed that all the stars in any given cluster are of similar age and at the same distance.  Therefore, a color-magnitude diagram provides the phase of evolution of any particular star and a clear distinction between Red Giant Branch (RGB) and AGB stars.  It is assumed that all stars will show similar abundance characteristics.  However, the globular cluster M15 has been shown to have a real star-to-star variation in the abundances of elements made by neutron capture (\citet{2005AJ....130.1177C} and references therein).  Three previous studies of the metal-rich cluster 47 Tucanae (\citet{1992AJ....104.1818B}, hereafter BW92, \citet{2004A&A...427..825J}, hereafter J04, and \citet{2005A&A...435..657A}, hereafter AB05) involved an extensive element abundance analysis of RGB stars.  Results of specific interest to this paper are shown in Table \ref{wall47tuc} with the number of stars observed in each study also listed.  One key result from this work was the large range in light s- and heavy s- process element abundances.  

\placetable{wall47tuc}

The AGB stars differ from their RGB progenitors as a result of subsequent nucleosynthesis and the TDU \citep{2003agbs.conf.....H}.  However, the AGB stars in 47 Tucanae are thought to have too low a mass to undergo the TDU phase responsible for producing s-process enhancement on the star's surface.  Consequently, no major s-process enhancement over their RGB progenitors is expected to be observed in these stars.  

Section \ref{obsandred} discusses the observations and the methods used for the reduction of the data.  Section \ref{analysis} describes the method used for analysis of the data, including stellar parameter determination, atomic and molecular line lists chosen and the method for establishing oscillator strengths.  The results of this research are presented in Section \ref{results} and trends in both light and heavy s-process elements are given along with the r-process element, Eu.  Finally, in Section \ref{Discussion} some possible explanations of these results are discussed.  Specific theories are presented along with justification and ideas for further work.

\section{Observations and Reduction}\label{obsandred}

Many photometric studies of 47 Tucanae have been undertaken, and the resulting color-magnitude diagrams provide an extensive list of potential AGB star candidates.  Specifically the color-magnitude diagrams produced using data from \citep{1978AJ.....83..376C} and the IRAF 2MASS study \citep{2006AJ....131.1163S}\footnote{http://irsa.ipac.caltech.edu}, and a spectroscopic study by Mallia (1978), have resulted in a list of AGB stars which were chosen for this work (Table \ref{atmosparam}).

\placefigure{hrd}

While Mallia classified these as AGB stars, other photometric data (\citet{1978AJ.....83..376C, 2006AJ....131.1163S}) confirmed AGB status for 5 of the stars but found that two stars, W66 and W68, fall on the RGB in the color-magnitude diagram (see Figure \ref{hrd}).  The classification from Figure \ref{hrd} was adopted, breaking our group into five AGB and two RGB stars.  All seven stars were observed using the 3.9m Anglo-Australian Telescope and UCLES during 2004 August.  Details of preciese coordinates of the target stars and recent magnitudes are shown in Table \ref{obsstars}.  Spectra were obtained for all wavelengths from 4400 to 6900 \AA.  The spectral resolution was about 40,000 with the EEV2 detector, with the slit width set to 1.5 arcsec.  In order to give complete coverage over the required wavelength interval, observations were centered around 5363\AA.  With all stars having magnitudes brighter than V=13.0, as many as five exposures were taken of each object, with an exposure time always less than 1800 seconds to minimize cosmic ray pollution.  The data were reduced at the University of Canterbury using the FIGARO software.  The reduced exposures for one star were then co-added to produce overall signal-to-noise ratios in the final spectrum of greater than 40.

\section{Analysis}\label{analysis}
 Atmospheric parameters were deduced using spectroscopic analysis of Fe I and Fe II lines.  Equivalent widths of Fe I and Fe II lines were used as input into the current version of the spectrum analysis code MOOG \citep{1973PhDT.......180S} and run with a variety of stellar atmosphere models until the most appropriate one was converged upon.  The relationship between derived abundance and excitation potential enabled the effective temperature to be determined, while the abundance and log(W/$\lambda$) relationship helped to refine the microturbulent velocity, see Figure \ref{param}.  The gravity of the model was established by comparing the abundance agreement of the Fe I and Fe II spectral lines.  

\placefigure{param}

The most acceptable stellar atmosphere was chosen to be the one with which the Fe I and Fe II abundances never differed by more than \underline{+}0.1 dex.  Table \ref{atmosparam} lists the atmospheric parameters settled on for each star analyzed.    

\placetable{atmosparam}

The effective temperatures and log g values calculated from photometry are also shown.  Photometric effective temperatures were found using the calibrations with colors and [Fe/H] published by \citet{1999A&AS..140..261A}, while photometric log g values are found via basic stellar structure equations.  In most cases, the temperatures found spectroscopically are $\sim$200 K greater than those found via the Alonso calibration. The J-K temperatures are consistently lower than the B-V temperatures by about 200K.  Reasons for this are unknown, although the Alonso (1999) paper claims that the established calibrations begin to diverge at T$<$4500K.  The spectroscopic and photometric gravities agree to within \underline{+}0.3 dex.

All stellar atmosphere models used for these stars were found via interpolation of the Bell et al.(1976) model grids for cool stars.  Determination of the correct stellar atmosphere model is crucial, as incorrect atmospheric parameters can result in large abundance discrepancies. 

Another crucial choice in the abundance analysis is the selection of spectral features and atomic line data.  Care was taken to select unblended lines wherever possible, although due to the presence of molecular bands blended lines are often unavoidable.  Over the wavelength range observed, 4400 to 6900\AA, molecular bands play a major role in stars at these temperatures, with the dominant feature being bands of TiO.  Molecular line parameters for TiO were taken from the lists published by Plez (1998).  

Approximate equivalent widths were measured and all lines with line strengths log (W/$\lambda$)$>$-4.5 were excluded from the final analysis.  A list of all lines and parameters used for the analysis of heavy elements is shown in Table \ref{linesused}.  All atomic line parameters were taken from Kurucz's lists \footnote{http://kurucz.harvard.edu}, as were molecular line parameters for MgH, CN and C$_{2}$.  However, for greater accuracy, oscillator strengths for the s- and r- process spectral features were refined from the original values.  A difference of \underline{+} 0.1 dex in log gf will effect the derived abundance by as much as \underline{+}0.1 dex, so it is crucial to define the log gf values as accurately as possible.  There are currently two different methods for establishing oscillator strength values.  Recent laboratory values have been published for Y \citep{1981ApJ...248..867B}, Zr \citep{2001ApJ...556..452L}, La \citep{2003ApJS..148..543D} and Nd \citep{1982ApJ...261..736H}, but it is also possible to determine values via a reverse solar analysis.  Hyperfine splitting was taken into account for all La II and Eu lines.  The effects of hyperfine splitting in lines of Nd II are minimal so was not included for the Nd II lines in this analysis.

In this research, the latter method was chosen and a reverse solar analysis was undertaken to establish log gf values of all heavy element lines.  A solar model of T$_{eff}$=5770K, log(g)=4.44 and v=1.4km s$^{-1}$ was used, with the published solar abundance values of \citet{1989GeCoA..53..197A}.  This method results in a solar model being built into the abundance analysis.  However, as is shown in Figure \ref{loggf} the values obtained via reverse solar analysis correlated well with recent lab values, thus minimizing any effect of incorporating the chosen solar model.  This agreement between lab and solar log gf values, as little as \underline{+}0.1 dex, ensures there is a minimal effect of using the solar model to define log gf values.

\placefigure{loggf}

\section{Results}\label{results}

Results for all the stars in our program are shown in Table \ref{47tucres}, which gives abundances for the s-process elements studied.  The standard notation is adopted, i.e. [X/Fe]=log(X/Fe)$_\star$-log(X/Fe)$_\sun$.  Mean values and associated $\sigma$ values are shown in Table \ref{sigmas}, along with ratios of [ls/Fe], [hs/Fe] and [hs/ls].  Mean values are shown for all seven stars observed, and a second mean value shown for the sample with the two RGB stars ignored.  In general, there is little difference between the mean values derived from the two samples.  For the purposes of this research `ls' is defined as the average of the abundances obtained from YI, YII, ZrI and ZrII, while `hs' is that obtained from LaII and NdII.   Sensitivity of abundances to the chosen model atmosphere parameters was undertaken over the range of derived atmospheric parameters, and results of this analysis are shown in Table \ref{deponparam}.  The analysis was undertaken using approximate equivalent widths for these lines, although due to the use of spectrum synthesis for s-process lines, these equivalent widths are indicative only.  Over this range of temperatures, the sensitivity of the derived abundances is small.  Sensitivities are also similar to previously published results for AGB stars of similar temperatures and gravities \citep{1985ApJ...294..326S}.

\placetable{47tucres}
\placetable{sigmas}
\placetable{deponparam}

\subsection{Fe Abundance}

The three previous studies of RGB stars used for comparison in this research found [Fe/H] to be between -0.81 \citep{1992AJ....104.1818B} to -0.67 \citep{2005A&A...435..657A}.  This research finds a mean [Fe/H] value of -0.60 \underline{+}0.20, in agreement with the more recent results of Alves-Brito et al.  However, the range of [Fe/H] in individual stars is between -0.55 and -0.71.  This mean [Fe/H] value falls within previously published values and agrees with the accepted value for the metal-rich 47 Tuc, [Fe/H]$\sim$-0.7 \citep{1996AJ....112.1487H}.  Fe I and Fe II abundance values agree to better than 0.1 dex.  The sigma value of the derived [Fe/H] value is $\sigma$=0.04.  It should be noted that both \citet{1992AJ....104.1818B} and \citet{2005A&A...435..657A} also found a range in Fe abundance in their studies of RGB stars, perhaps suggesting that a non-uniform Fe abundance in 47 Tuc is present.

\subsection{Light s-process}
As can be seen from Table \ref{47tucres}, all seven stars analyzed showed marked s-process enhancements relative to Fe.  Examples of spectrum synthesis of Y and Zr are shown in Figures \ref{specy} and \ref{speczr}.  The observed spectrum of the star W66 is shown as crosses.  The four lines show synthetic spectra with different abundance values.  The spectrum was always synthesized with the particular species absent from the synthesis to ensure that the line analyzed was not overly blended.  This synthesis is shown by the short dashed line.  The other three lines show syntheses with varying enhancements.

\placefigure{specy}
\placefigure{speczr}

It was possible to measure both neutral and ionized lines for the light s-process components, Y and Zr.  Generally, the two ionization levels within a given element agreed (see Figure \ref{iiicompare}), although two stars, W68 and 384, showed discrepancies between Zr I and Zr II of over 0.2 dex.  The possibility of this being due to an incorrect choice of gravity in the stellar model seems slim.  However, the cause of this disagreement between ZrI and ZrII is uncertain.  As is seen in Figure \ref{iiicompare} the Y I and Y II abundance and Fe I and Fe II abundance both agree to $<$0.05 dex in both cases. 

\placefigure{iiicompare}
 
From Table \ref{wall47tuc}, it is clear that in the RGB stars of 47 Tuc all previous studies found an underabundance of the s-process element Zr.  The average derived [Zr/Fe] value is $\sim$-0.3.  However, this study has found that all the giant stars observed (both AGB and RGB) gave an average [Zr/Fe] value of $\sim$+0.6.  This is a major difference from the value found in the previous studies of the RGB stars of the same cluster, so further investigation was undertaken.  In order to reliably compare between the abundances derived from the RGB stars and this paper, a check was undertaken to ensure the results were consistent and no external discrepancies may be causing this large difference.  

The equivalent width data from BW92 was obtained from the publishers and used as input into MOOG.  This ensured that any potential error from the modeling or synthesis was minimized.  When the equivalent width data from BW92 was used in MOOG, it produced the derived abundances that BW92 reported.  This demonstrates that any source of error from the choice of models or synthesis technique is negligible.  A second check was undertaken in which the published Zr abundance from BW92 was synthesized and compared with the observed spectrum of one of the RGB stars from this study (W66).  From Figure \ref{zrcomp} it is clear that there is an enhancement in the Zr abundance relative to the BW92 RGB values.  It should be noted that \citet{2002AJ....123.3277R} found a similar range of Zr abundances in bright giant stars in the globular cluster M71.

\placefigure{zrcomp}

\subsection{Heavy s-process}

It was only possible to measure ionized lines for the heavy s-process components, La and Nd.  Ba was not used as many Ba lines are saturated and consequently are too strong to produce reliable abundance results, with log (W/$\lambda$) values $>$-4.0.

From Table \ref{wall47tuc}, it is clear that in the RGB stars of 47 Tuc, BW92 found a slight enhancement of La II, $\sim$+0.2 dex while AB05 found slight enhancements in both La, $\sim$+0.1 dex, and Ba, $\sim$+0.3 dex.  The only element studied in common with this study was La II, which was found to be enhanced by between $\sim$+0.2 and $\sim$+0.5 dex in both the RGB and AGB program stars of this study, with a mean value of +0.31.  This is clearly a larger enhancement than that previously found in the RGB stars of earlier studies.  Nd II, which is also a heavy s-process element, from the same peak as La and Ba, was also found to be enhanced by about the same amount, with a mean value of +0.42 dex.  More interestingly, both elements appear to be similarly enhanced in individual stars, suggesting that the origin of this enhancement is increasing both heavy s- elements together.

\subsection{r-process (Eu)}
The Eu abundance is of interest as Eu is a solely r-process element and, consequently, is a good indication of the extent of r-processing in the primordial material of this cluster.  An example of the spectrum synthesis on the Eu 6645\AA line is shown in Figure \ref{speceu}.

\placefigure{speceu}

All previous studies of the RGB stars found Eu to be well enhanced, with values of [Eu/Fe] ranging from +0.17 to +0.33.  This study also found Eu to be slightly enhanced, with values ranging from around solar, $\sim$-0.08, to $\sim$+0.26.  This general agreement with the RGB stars lends strong support to claims of a genuine s-process enhancement as it is not expected that stars in a globular cluster should differ in their r-process abundances.

\subsection{Na Abundance}\label{na}

The Na abundance was also measured as a way of estimating CN strength.  \citet{1981ApJ...245L..79C} and \citet{1979ApJ...230L.179N} observed a distinct CN bimodality in 47 Tuc, with the added feature that Na was enhanced in the CN strong stars.  There is also a reliable correlation between enhanced Na and CN strength.  Na abundances in these program stars varied from +0.3 to +1.0.  This may be an indication of a range of CN strengths in these stars.  When comparing the derived Na abundance to the CN strength derived from the CN feature at 4100\AA published by \citet{1978A&A....70..115M}, there appears to be only slight agreement.  The significance of the CN strength will be discussed further in Section \ref{mydiscuss}.

\section{Discussion}\label{Discussion}

The results obtained in this work need to be interpreted in terms of our understanding of the AGB phase of evolution and binary star evolution, as well as the possibility of intrinsic star to star variations within the cluster and possibly as a result of a previous generation of stellar processing.  The main reason the AGB phase is so interesting to
astrophysicists is the nucleosynthesis and heavy element formation that can occur
throughout this phase. 

\subsection{s-process element formation in AGB stars}\label{SAGB}

The s-process consists of neutron capture followed by a beta decay, where the neutron capture is slow relative to
the beta decay.  Two
neutron sources have been
suggested: $^{22}$Ne($\alpha,n$)$^{25}$Mg and $^{13}$C$(\alpha,n)^{16}$O.  If the temperature
is sufficiently high the $^{22}$Ne reaction 
provides the neutrons.  However, this is achieved only in relatively massive AGB stars (M$>$3M$_\sun$) when the temperature in the helium-rich intershell region reaches 10$^{8}$K.  This temperature can be reached in stars which have masses around M$\sim$1-2M$_\sun$ but usually during the later thermal pulses.  The second possible  neutron source is the {$^{13}$C reaction.  During the dredge-up phase (TDU) of the thermal pulse, partial mixing occurs
at the bottom of the convective envelope and protons are deposited into the
intershell region.  The protons then combine with the $^{12}$C present 
to form $^{13}$C.  If there are enough protons in the envelope
then the CN cycle can proceed, forming $^{14}$N.  One of the consequences 
of this process is a layer of enhanced $^{13}$C which forms in a pocket, 
known as the $^{13}$C pocket.  This $^{13}$C pocket is where the
majority of s-process element formation is proposed to occur.  

During the next thermal pulse, these s-process products are mixed into
the convective shell, where further nucleosynthesis can occur, including a brief burst of neutrons from the $^{22}$Ne source, if the temperature is high enough.  After this thermal pulse, the products
formed in this convective shell are fully mixed to the surface of
the star during the dredge-up phase (see \citet{2001MmSAI..72..255L} and references therein). 

Both the first and second dredge up act to mix to the
surface regions that have undergone H burning through the CNO cycles
and consequently increase the abundances of $^{13}$C, $^{14}$N and
$^{4}$He.  There are also associated decreases in $^{12}$C, $^{16}$O
and $^{18}$O.  The minor difference between the two is that the second
dredge up mixes material that has burned all of the available H and therefore
produces large abundance changes, whereas the first dredge up mixes
regions that have only undergone partial H burning.

\subsection{Specific discussion of these results} \label{mydiscuss}

There are three possible scenarios to be considered when explaining the observed enhancements of s-process elements in these giant branch stars.  First, a possible adjustment may be needed in low-mass AGB modeling (\ref{AGBmodel}); second, these AGB stars could be the by-product of an earlier evolved binary system (\ref{binary}); or third, it may be that the observed enhancement is a result of a genuine star to star overabundance present in 47 Tuc (\ref{scatter}). 

\subsubsection{AGB modeling considerations}\label{AGBmodel}

After core He exhaustion a star moves toward the AGB.  The first part of this evolution is called the Early-AGB, which is terminated when the first thermal pulse occurs and the star starts its Thermally-Pulsing AGB evolution.  Following most thermal pulses, dredge-up will mix to the surface any s-process elements in the intershell region.  With each subsequent pulse the overall neutron exposure increase and hence the relative abundances of the ls (eg. Sr, Y and Zr) and hs (eg. Ba, La and Nd) elements change.  During the earlier pulses we expect the ls elements to dominate but this will shift to the hs element as neutron capture progresses.  The number of neutrons released are expected to be largely independent of the metallicity of the host star, being produced mostly by some mixing or hydrodynamic process.  Hence, if the number of neutrons is primary, then the neutron exposure per seed will depend mostly on the number of seeds present; thus stars with low [Fe/H] will absorb more neutrons per seed than stars with a higher [Fe/H].  Thus we expect low metallicity stars to show relatively more hs species while higher metallicity stars, such as found in 47 Tuc, would show more ls species (see Figure 12 in Busso et al. 1999).

The ls values of the five AGB stars in this study range from [ls/Fe] = +0.36 to +0.84, while the hs values range from [hs/Fe] = +0.26 to +0.51.  The corresponding [hs/ls] value gives an indication of the relative strengths of these two peaks and ranges from [hs/ls] = -0.09 to -0.40.

Recent theoretical predictions of ls and hs abundances as a function of [Fe/H] are presented in \citet{2004MmSAI..75..700G}.  As is discussed in that paper, at any given metallicity, a large range of $^{13}$C pocket efficiencies needs to be adopted to incorporate the large scatter in the observational data.  Stellar models predict a large range of [ls/Fe], [hs/Fe] and [hs/ls] depending on the choice of the $^{13}$C pocket parameterization.  The s-process enhancement varies over the first few thermal pulses, reaching an asymptotic value after about 10 TDU episodes.  

Figures \ref{fels}, \ref{fehs} and \ref{fehsls} show the predicted [ls/Fe], [hs/Fe] and [hs/ls] versus [Fe/H] respectively.  The lowest mass predictions made are for a 1.5M$_\sun$ stellar model and various $^{13}$C pocket efficiencies, varying from the standard choice (ST), M($^{13}$C)=3 x 10$^{-6}$ M$_\sun$ burnt per pulse, to both higher and lower values.  A major conclusion to be drawn from these figures is the value chosen for the $^{13}$C pocket parameterization.  Figure \ref{fels} suggests a choice of between ST/6 and ST/12 and Figure \ref{fehs} suggests between ST/3 and ST/6.  Figure \ref{fehsls} presents an ambiguity of parametrization choices, however, the values of ST/6 to ST/12 are not ruled out by these [hs/ls] values, with ST/12 remaining a viable option.  While the specific values vary, all three figures agree that the choice for the $^{13}$C pocket efficiency should be less than the standard choice.  These results may also be explained by the amount of $^{12}$C that is left in the He intershell.  This is less than is present in the $^{13}$C pocket, and if it is comparable to the ST/10 value then this may be the origin of the neutron source.  

Taken at face value this scenario would be a possible explanation of the observations even though TDU has not been predicted to occur in such low mass stars and at such low luminosities.  However, one major difficulty with this is that the two RGB stars in this study have similar abundances to the AGB stars and already have ls and hs enhancements.  Consequently, little or no TDU mixing would be required to transform the RGB ls and hs element abundances to the AGB abundances for these elements.  

\subsubsection{Binary star system}\label{binary}

The typical mass of a genuine AGB star in 47 Tuc is $\sim$0.75M$_\sun$, slightly lower than the turn-off mass of M$\sim$0.85M$_\sun$ \citep{1998ApJ...507..818G}, and thought to be too low to undergo TDU.  However, 47 Tuc has been shown to have a large number of blue straggler stars (BSS) \citep{2003ApJ...588..464F}.  A model has been suggested (Cristallo, private communication) in which a 0.7M$_\sun$ AGB star is formed through mass transfer in a binary star system.  In this scenario, two stars of initial mass 1.4M$_\sun$ and 0.5M$_\sun$ are in a binary system.  The larger star evolves in 2.5Gyr to the AGB and is massive enough to undergo s-process formation and the required TDU to enhance the surface abundance.  Once this star has filled its Roche lobe, this mass is transferred to the smaller companion, which begins life again near the zero-age main-sequence for a star with its new mass.  This new star, of M=1.1M$_\sun$ evolves over 8 Gyr to form an AGB star of 0.8M$_\sun$, enhanced in s-process from the previous evolution.  It is this AGB star that we observe today.  This model is entirely consistent with the accepted age of 47 Tuc of 11.1 \underline{+}1.1 Gyr.  However, it seems unlikely that all seven of the stars we observed have undergone this scenario.

\subsubsection{Genuine star-to-star scatter}\label{scatter}

Figure \ref{fehhsls} presents a third explanation.  If we accept the placement of W66 and W68 in the CMD, these two stars appear to lie on the RGB of the CMD.  While Mallia (1978) claimed these were AGB stars, two specific photometric studies \citep{1978AJ.....83..376C, 2006AJ....131.1163S} give values of V and B-V  and J and J-K respectively that place these stars on the RGB.  If this can be believed and these stars are on the RGB then their values, when taken with the values given by previous studies, indicates a true spread in RGB values of [hs/ls].  This may be indicative of the presence of a broad distribution of [hs/ls] in 47 Tuc.  The other alternative is that there are zero point problems between the different studies of RGB stars in 47 Tuc.  Only a larger, consistent survey using medium to high resolution data can resolve this issue.  

Figures \ref{ywitht} and \ref{zrwitht} show two spectra in the region of light s-process elements stacked in order of increasing temperature.  From these figures it is clear that there is no real trend with effective temperature, suggesting that any spread of s-process enhancements are genuine and indicative of a real star-to-star scatter present in the stars of 47 Tuc.  This can also be seen when considering the spread of RGB abundances shown in Figure \ref{fehhsls}.  

Further strengthening the argument for genuine star-to-star scatter in abundances is the distinct spread in Na abundance we observed.  There is a well defined correlation between Na abundance and CN strength, as well as a firmly established bimodal CN strength in 47 Tuc \citep{1979ApJ...230L.179N}. There is a strong suggestion that this CN strength is indicative of the presence of two separate stellar populations in 47 Tuc.  The presence of at least two populations of stars in 47 Tuc may also explain the spread in [hs/ls] seen in Figure \ref{fehhsls} and would help to explain the range seen in the RGB stars.  This group of giant stars displays a Na abundance spread from [Na/Fe] = +0.31 to +1.04.  If this Na abundance spread is indeed evidence of CN strength, then it can be assumed that there is also a distinct CN spread in these giant stars.

\section{Conclusion}

AGB nucleosynthesis is dependent on a number of factors.  From the theoretical viewpoint, the most important are typically the stellar mass and composition and the mass loss rate during the later stages of their evolution.  Significant uncertainty exists at present concerning the efficiency of the third dredge-up phenomenon.  Details of the $^{13}$C pocket, both its formation and the resulting abundance profile, are still not known.  Much evidence exists to support the idea that a spread in $^{13}$C pockets is required to match the observed s-process abundances, both in stars and in pre-solar grains.  The details of the mass-loss prescription are also important: at what stage does the envelope lose its mass and at what rate?  This affects the evolution as well as the nucleosynthesis.   

Extensive observations of AGB stars are needed to help limit and refine the basic parameters assumed during AGB nucleosynthesis modeling.  The giant stars observed in this research show marked s-process enhancement relative to the RGB stars in other studies of 47 Tuc.  This is completely contradictory to the theoretical predictions since these AGB stars are believed to be too low in mass to undergo TDU and the related s-process element enhancements.  

Of the possible scenarios proposed to explain these enhancements, the most likely seems to be the possibility of a genuine spread in element abundances in individual stars in 47 Tuc.  Any star-to-star scatter would be explained by primordial processes, which would remove the need to reassess the assumptions made in low-mass AGB modeling.  A large sample survey of stars at various evolutionary phases and extensive abundance analysis of a range of key nucleosynthetic indicators would help to confirm whether this proposed star-to-star scatter is real.

\acknowledgments

ECW acknowledges the support of a University of Canterbury Research Award, a Departmental Scholarship and the Dennis William Moore Scholarship, all held for the duration of this research.  Thanks to Oscar Straniero and Sergio Cristallo for discussion regarding AGB evolutionary modeling and possible scenarios to explain the observed enhancements.  This research was partly supported by the Australian Research Council.  This research has made use of the SIMBAD database, operated at the Centre de Don\'{e}es Astronomiques de Strasbourg (CDS), France.  This publication makes use of data products from the Two Micron All Sky Survey, which is a joint project of the University of Massachusetts and hte Infrared Processing and Analysis Center/California Institute of Technology, funded by the National Aeronautics and Space Administration and the National Science Foundation.

Facilities: \facility{AAT}(UCLES)

%% thebibliography produces citations in the text using \bibitem-\cite
%% cross-referencing. Each reference is preceded by a
%% \bibitem command that defines in curly braces the KEY that corresponds
%% to the KEY in the \cite commands (see the first section above).
%% Make sure that you provide a unique KEY for every \bibitem or else the
%% paper will not LaTeX. The square brackets should contain
%% the citation text that LaTeX will insert in
%% place of the \cite commands.

%% We have used macros to produce journal name abbreviations.
%% AASTeX provides a number of these for the more frequently-cited journals.
%% See the Author Guide for a list of them.

%% Note that the style of the \bibitem labels (in []) is slightly
%% different from previous examples.  The natbib system solves a host
%% of citation expression problems, but it is necessary to clearly
%% delimit the year from the author name used in the citation.
%% See the natbib documentation for more details and options.

\clearpage

\begin{figure}
\epsscale{0.5}
\plotone{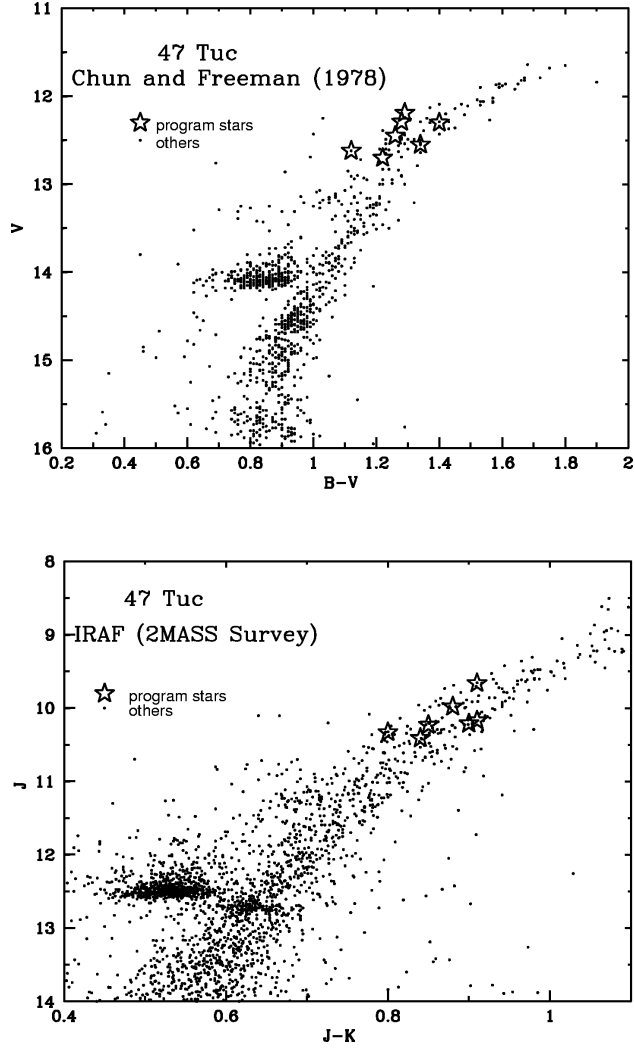}
\caption{Color-magnitude diagram of 47 Tuc with observed stars shown.  V and B-V values of most stars are taken from \citet{1978AJ.....83..376C}; the exception is W68, where the values are taken from \citet{1961ApJ...133..430W}.  The J and J-K values for all stars are taken from the IRAF 2MASS survey \citep{2006AJ....131.1163S}(http://irsa.ipac.caltech.edu).} \label{hrd}
\end{figure}

\clearpage

\begin{figure}
\epsscale{0.8}
\plotone{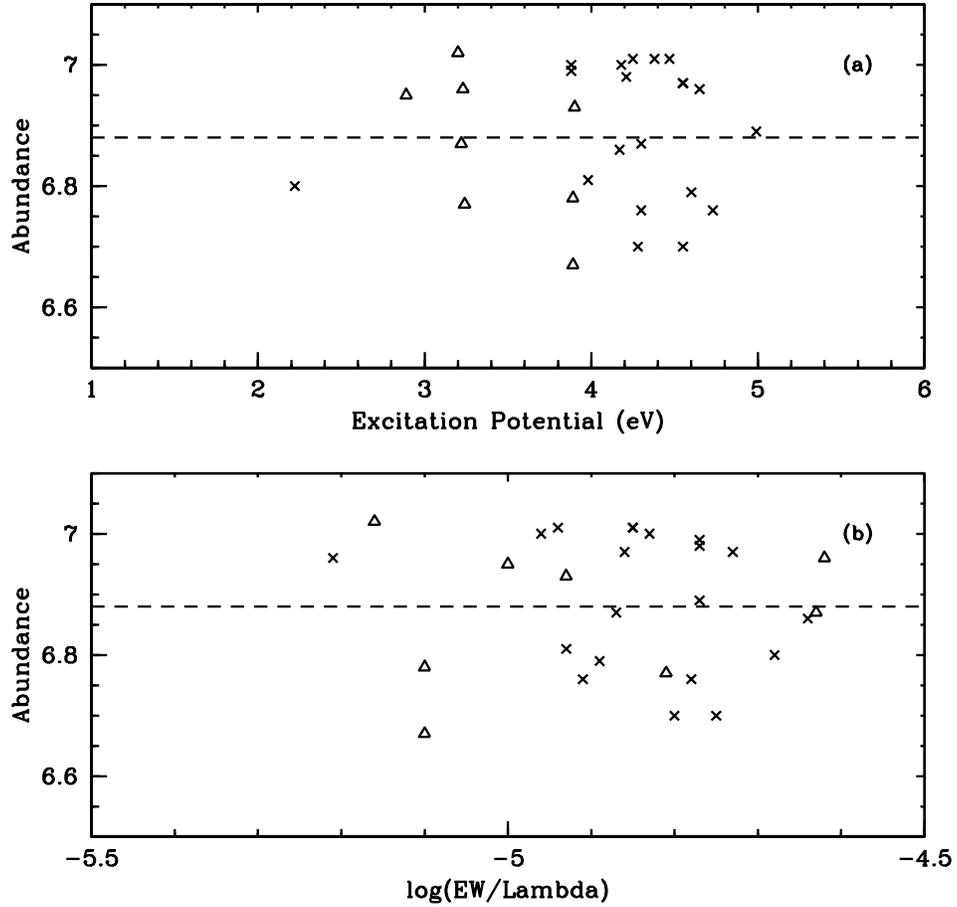}
\caption{Equivalent width analysis to refine atmospheric parameters for stellar models. The trend of excitation potential vs abundance (a) gives the effective temperature while the trend of log(EW/$\lambda$) vs abundance (b) gives the microturbulence.  The gravity if refined via comparison of the neutral (crosses) and ionized (triangles) species of Fe. \label{param}}
\end{figure}

\clearpage

\begin{figure}
\epsscale{0.8}
\plotone{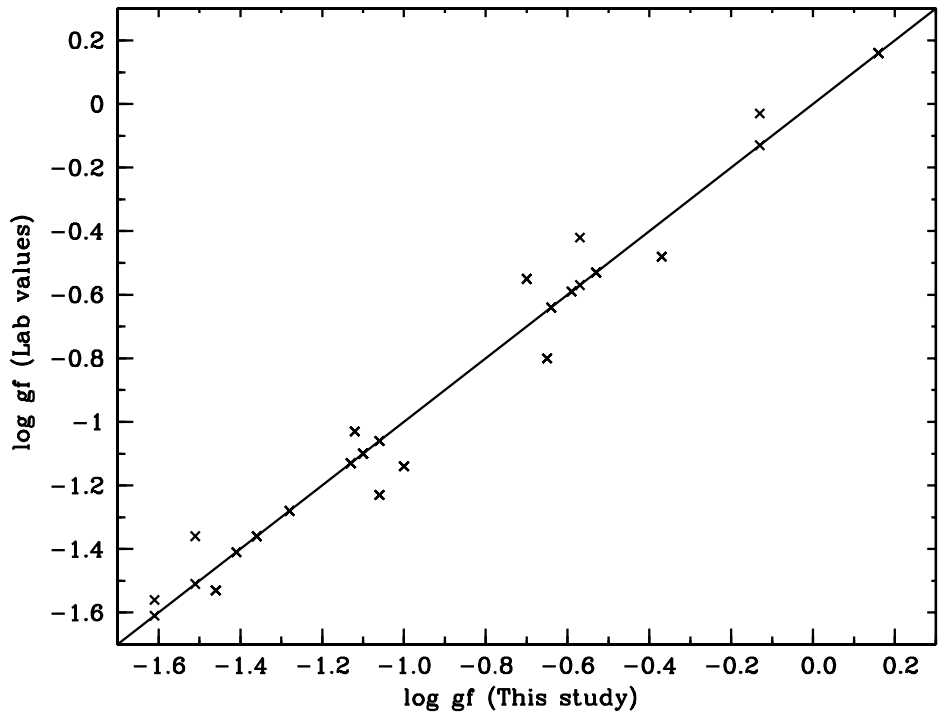}
\caption{Comparison of log gf values obtained via reverse solar analysis with laboratory values published previously (Y: \citet{1981ApJ...248..867B}, Zr: \citet{2001ApJ...556..452L}, La: \citet{2003ApJS..148..543D} and Nd: \citet{1982ApJ...261..736H}).  The straight line shows a perfect agreement.   \label{loggf}}
\end{figure}

\clearpage

\begin{figure}
\epsscale{0.8}
\plotone{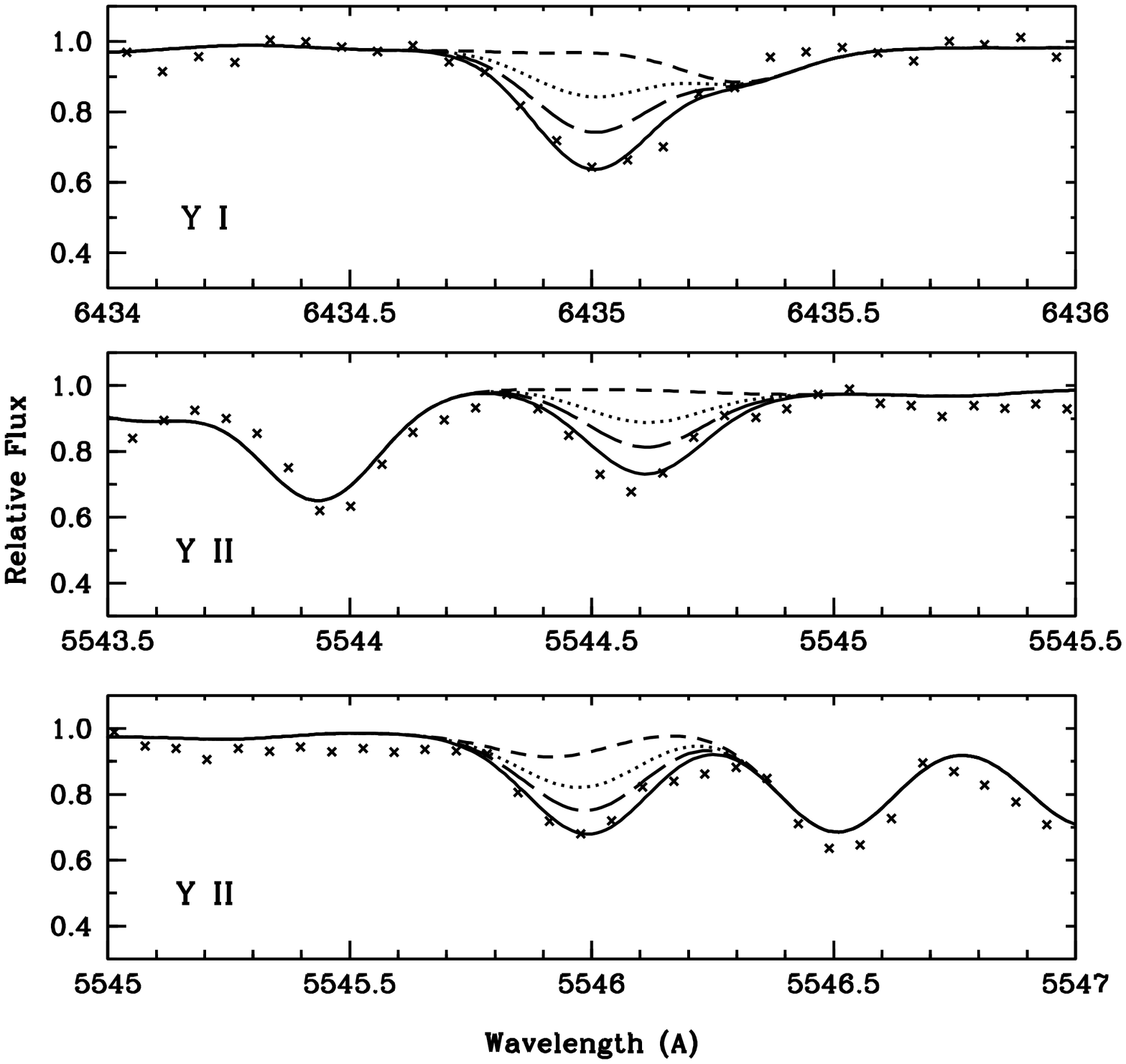}
\caption{Synthesis of Y I and II lines showing the best fit to the observed spectrum of the RGB star, W66.  The short dashed line shows spectrum synthesis with no Y present, while the other three lines show various Y enhancements.  The dotted line represents [Y/Fe]=0.0, the long dashed line represents [Y/Fe]=+0.3 and the solid line represents the best fit, [Y/Fe]$\sim$+0.67, to the observed spectrum (crosses).  \label{specy}}
\end{figure}

\clearpage

\begin{figure}
\epsscale{0.8}
\plotone{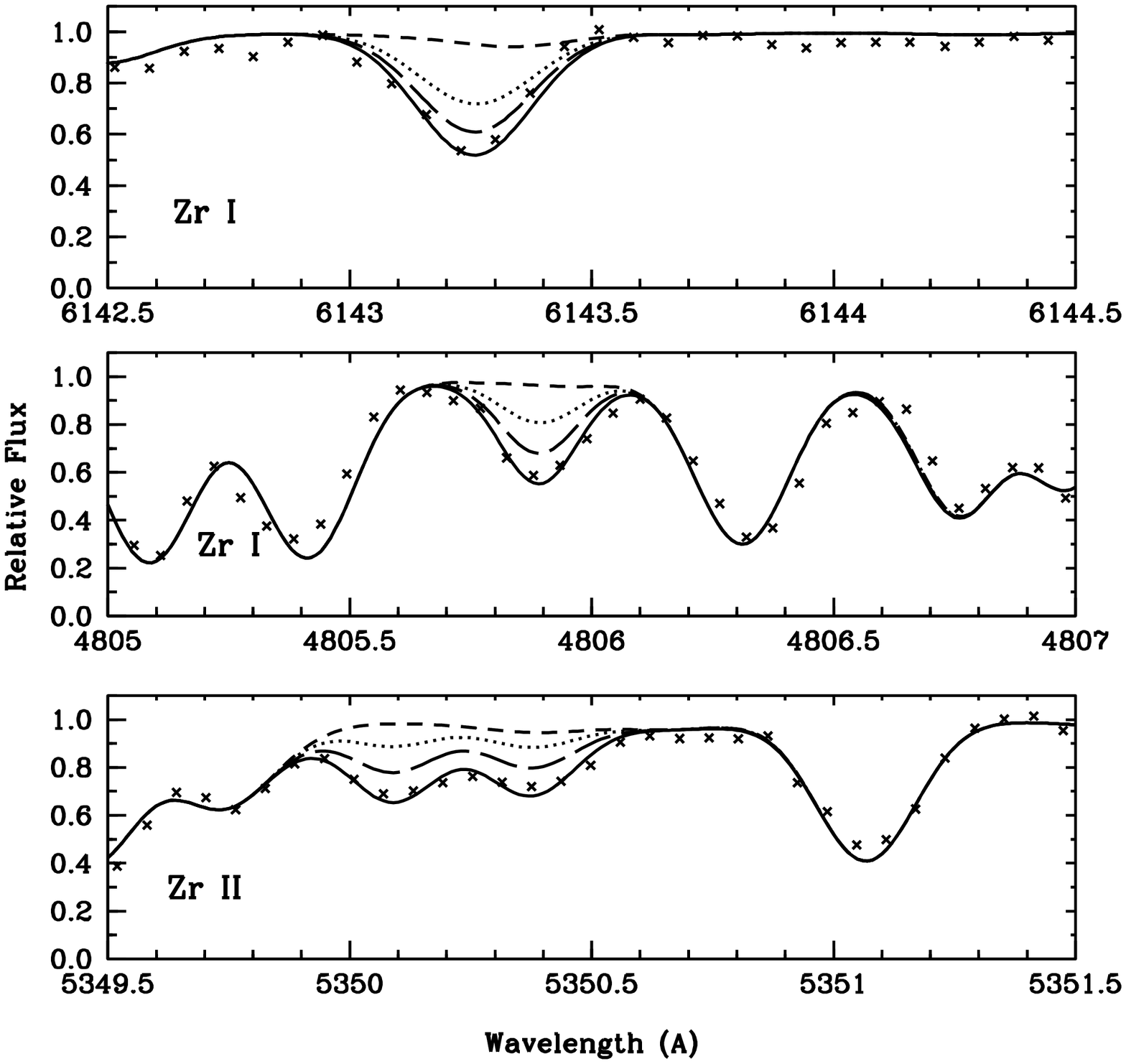}
\caption{Synthesis of three Zr I lines showing the best fit to the observed spectrum of the RGB star, W66.  The short dashed line shows spectrum synthesis with no Zr present, while the other three lines show various Zr enhancements.  The dotted line represents [Zr/Fe]=0.0, the long dashed line represents [Zr/Fe]=+0.3 and the solid line represents the best fit, [Zr/Fe]$\sim$+0.58, to the observed spectrum (crosses).   \label{speczr}}
\end{figure}

\clearpage

\begin{figure}
\epsscale{1.0}
\plotone{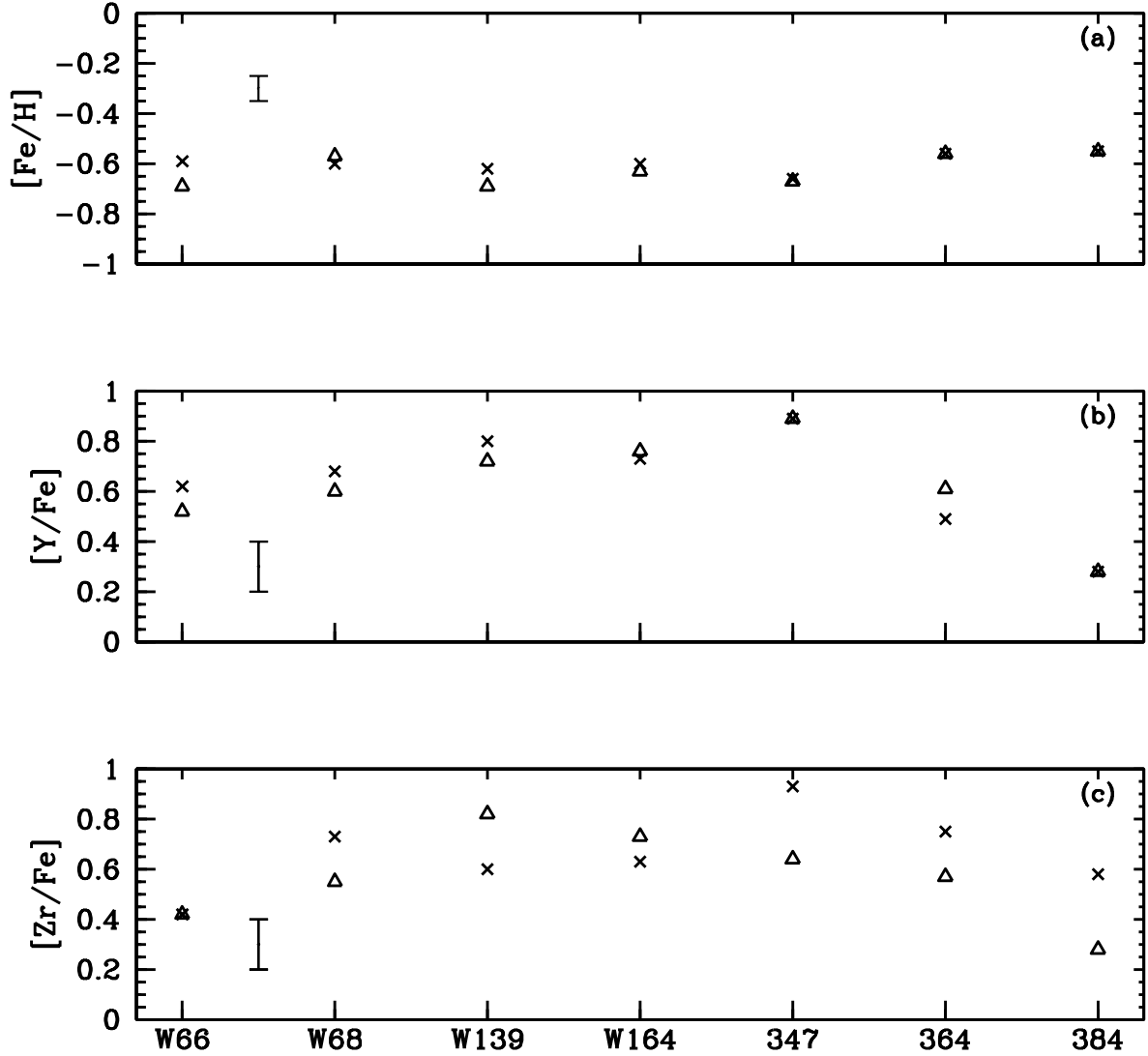}
\caption{Abundances of Fe (a), Y (b) and Zr (c), comparing derived abundances from neutral species (crosses) and ionized species (triangles), with applicable sigma values.  It is clear that both species agree to within the uncertainty shown.  \label{iiicompare}}
\end{figure}

\clearpage

\begin{figure}
\epsscale{0.8}
\plotone{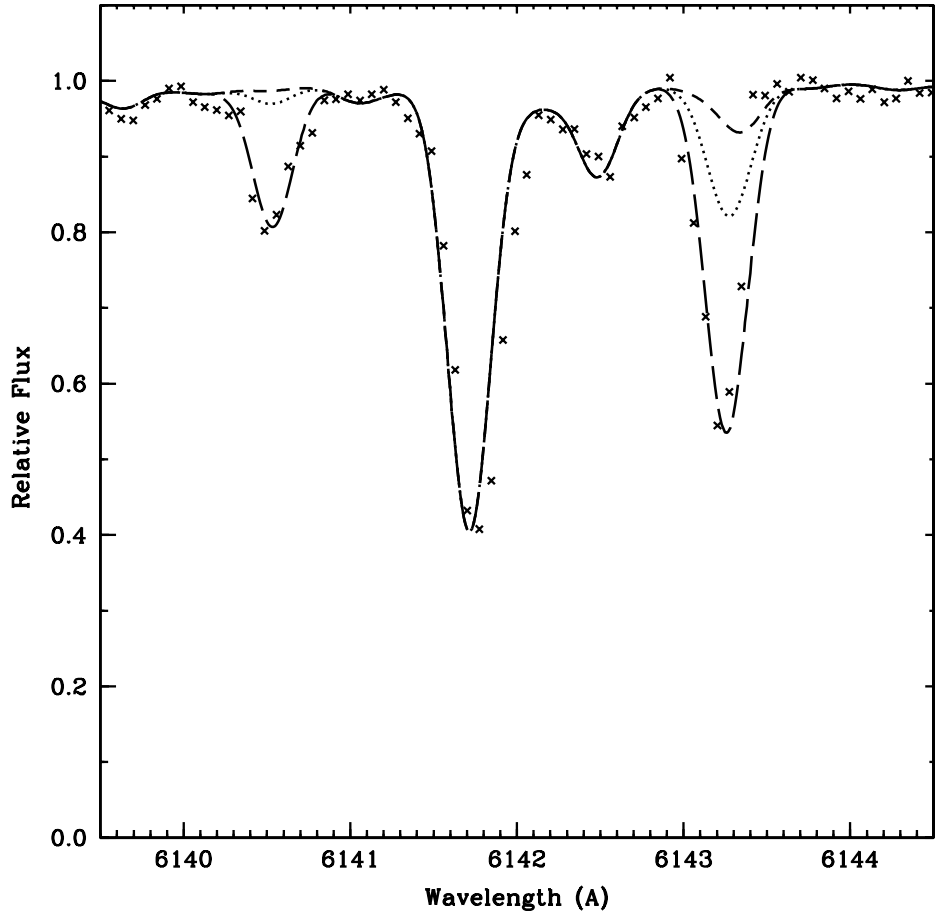}
\caption{Synthesis of Zr I 6140 and 6143\AA\ lines.  The crosses show the observed spectrum in the 47 Tuc RGB star, W66.  The short dashed line shows spectrum synthesis with no Zr present; the dotted line with the BW92 value of log $\epsilon$ $\sim$1.70 and the long dashed line with the best fit value of log $\epsilon \sim$ 2.90.  There is clearly an enhancement of Zr in this AGB star compared with the value found by BW92 in the RGB stars.  \label{zrcomp}}
\end{figure}

\begin{figure}
\epsscale{0.8}
\plotone{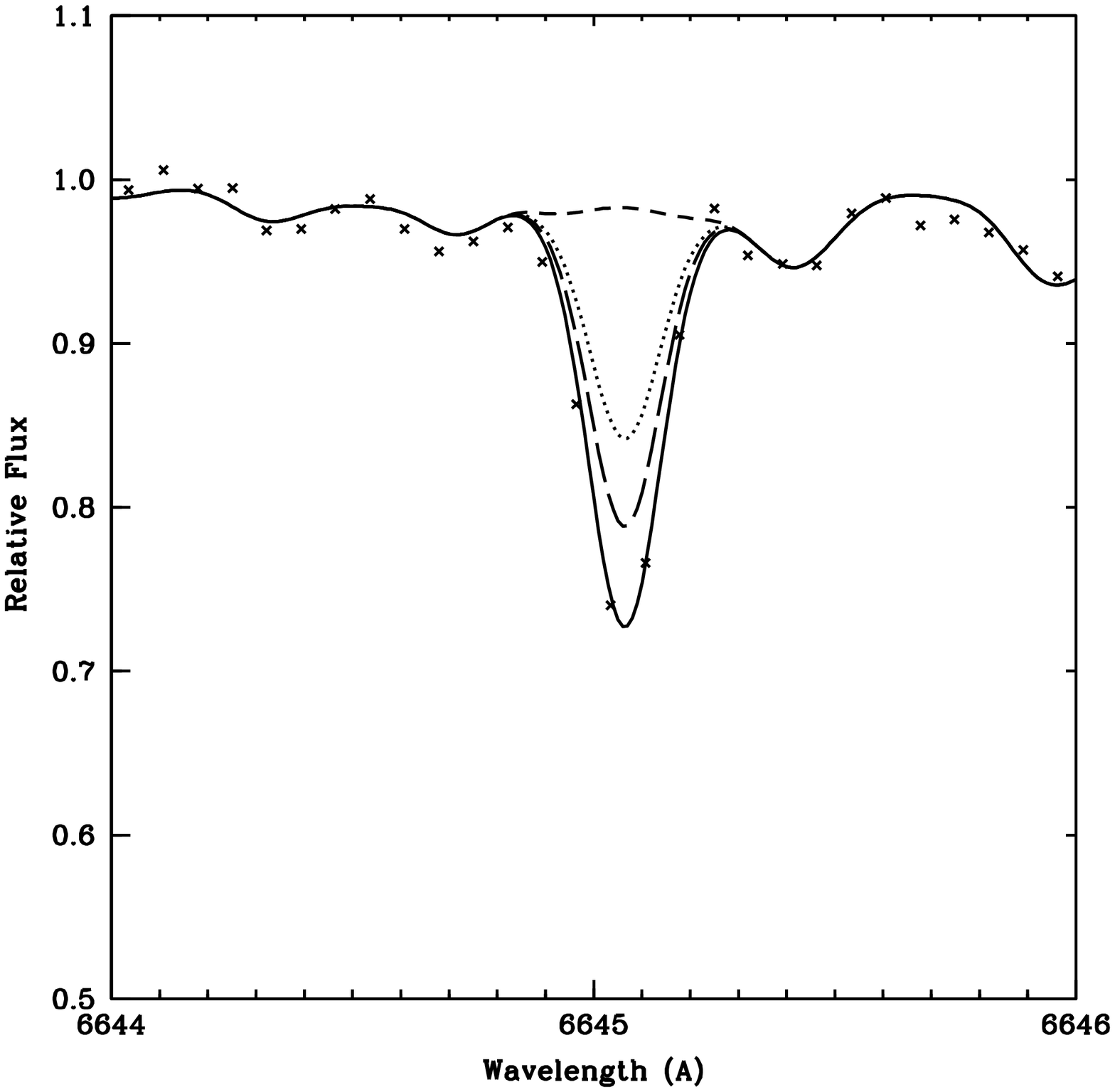}
\caption{Synthesis of Eu II 6645\AA\ line showing the best fit to the observed spectrum of the RGB star, W66.  The short dashed line shows spectrum synthesis with no Eu present, while the other three lines show various Eu enhancements.  The dotted line represents [Eu/Fe]=0.0, the long dashed line represents [Eu/Fe]=+0.15 and the solid line represents the best fit, [Eu/Fe]$\sim$+0.27, to the observed spectrum (crosses).   \label{speceu}}
\end{figure}

\clearpage

\begin{figure}
\epsscale{0.8}
\plotone{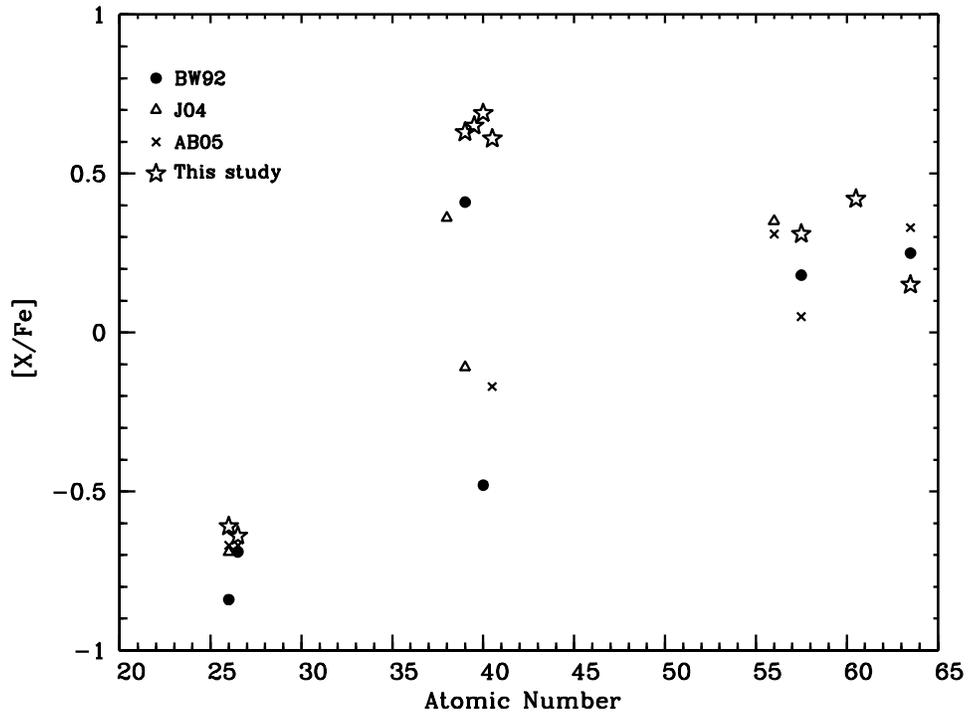}
\caption{Heavy element abundances in RGB and AGB stars of 47 Tuc. While the Fe abundance agrees within uncertainties, there is an obvious discrepancy between the RGB stars studied previously and the giant stars from this study in the s-process elements, particularly at the light s- peak.  \label{wallvswy}}
\end{figure}

\clearpage

\begin{figure}
\epsscale{0.8}
\plotone{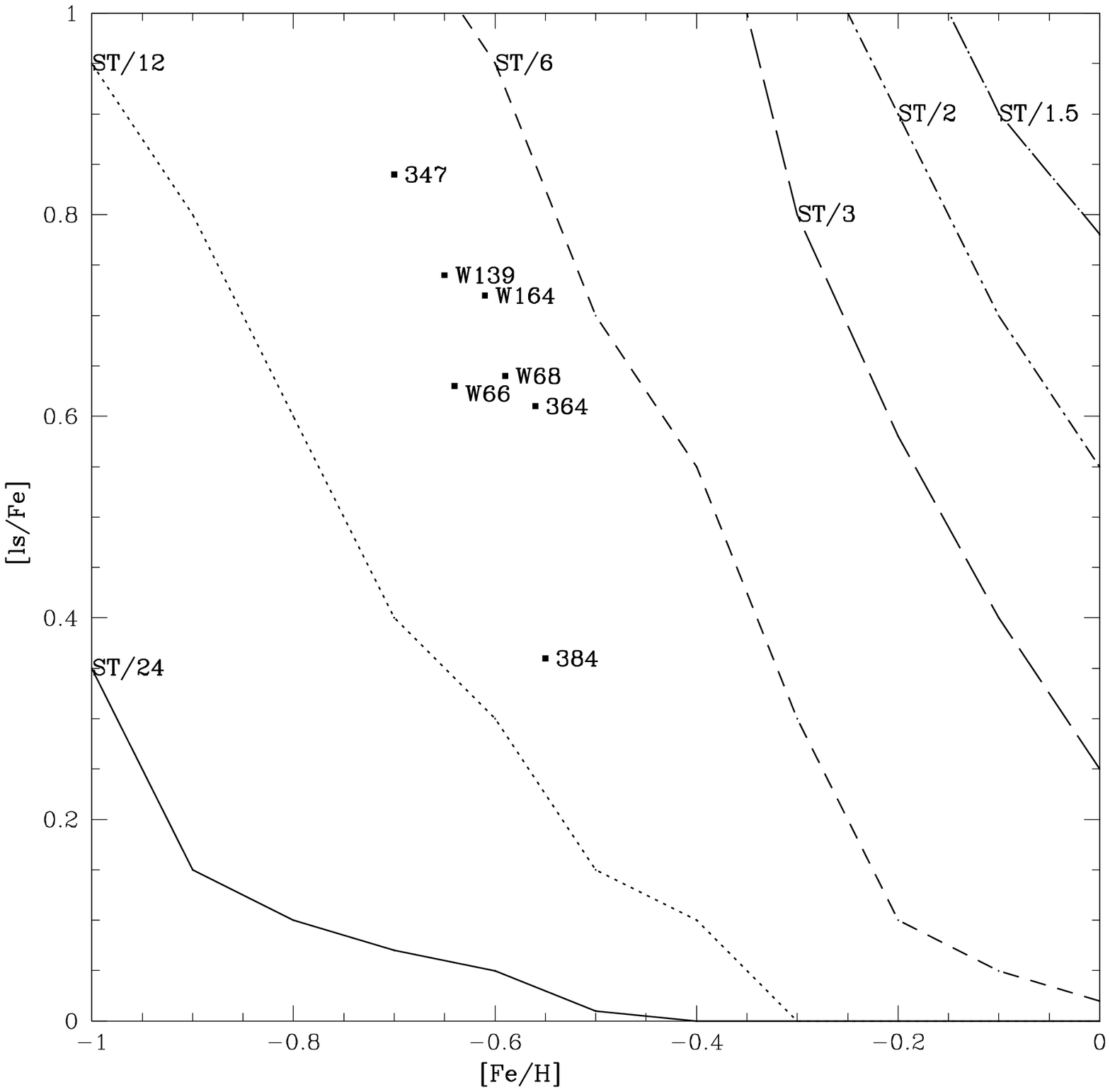}
\caption{Ratio [ls/Fe] with different choices of $^{13}$C pocket parameterizations. \label{fels}}
\end{figure}

\clearpage

\begin{figure}
\epsscale{0.8}
\plotone{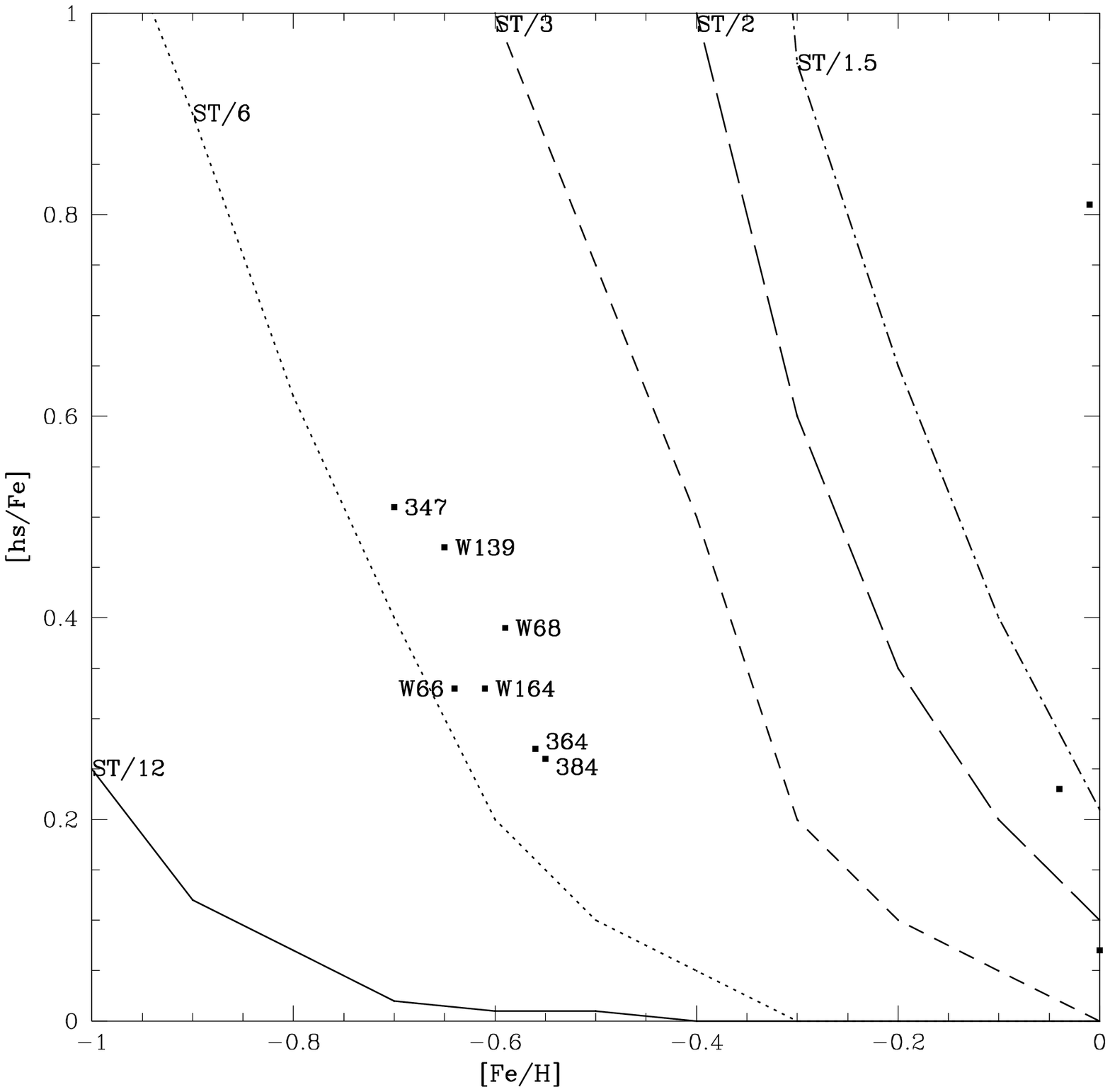}
\caption{Ratio [hs/Fe] with different choices of $^{13}$C pocket parameterizations.\label{fehs}}
\end{figure}

\clearpage

\begin{figure}
\epsscale{0.8}
\plotone{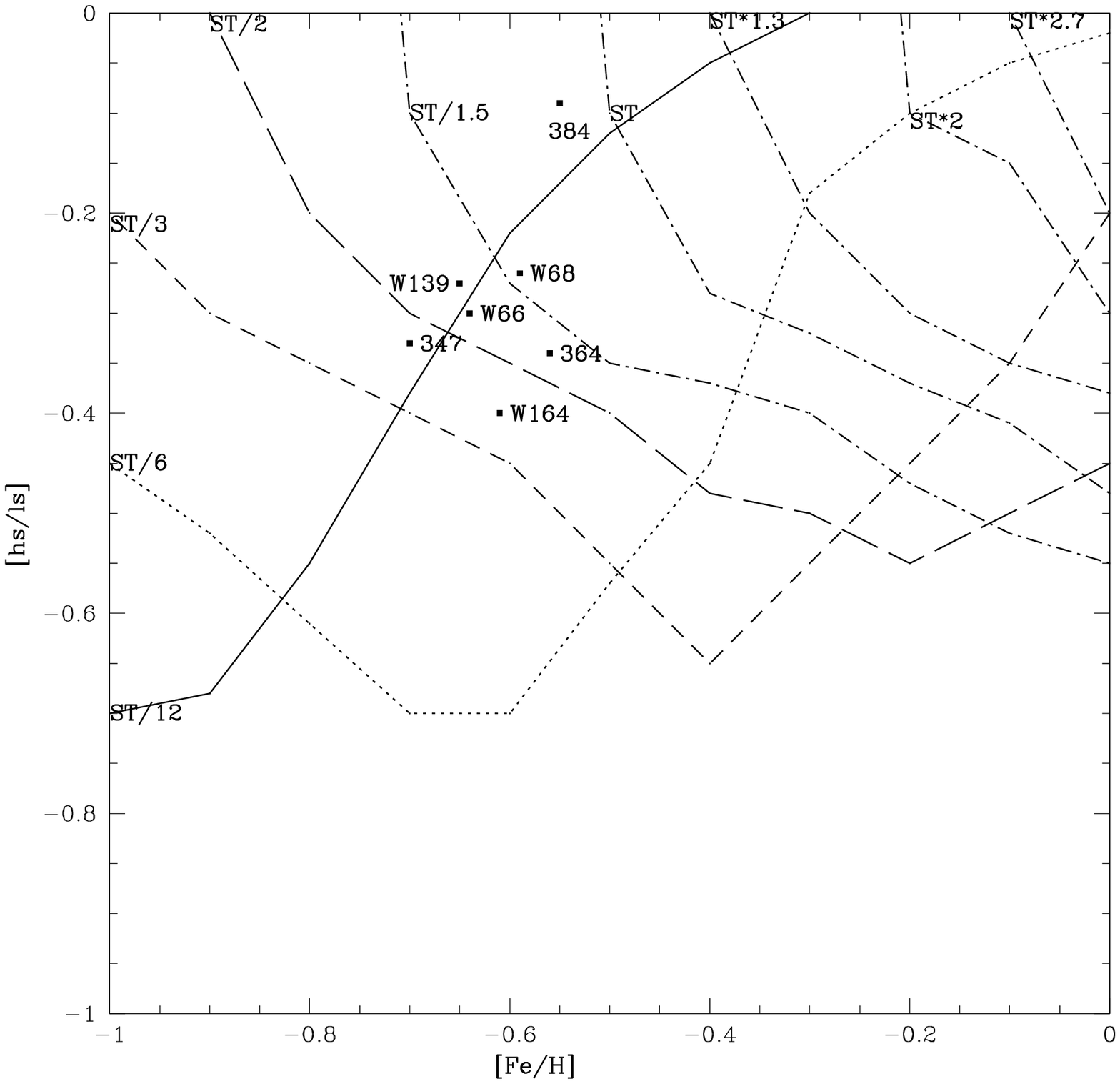}
\caption{Ratio [hs/ls] with different choices of $^{13}$C pocket parameterizations. \label{fehsls}}
\end{figure}

\clearpage

\begin{figure}
\epsscale{0.8}
\plotone{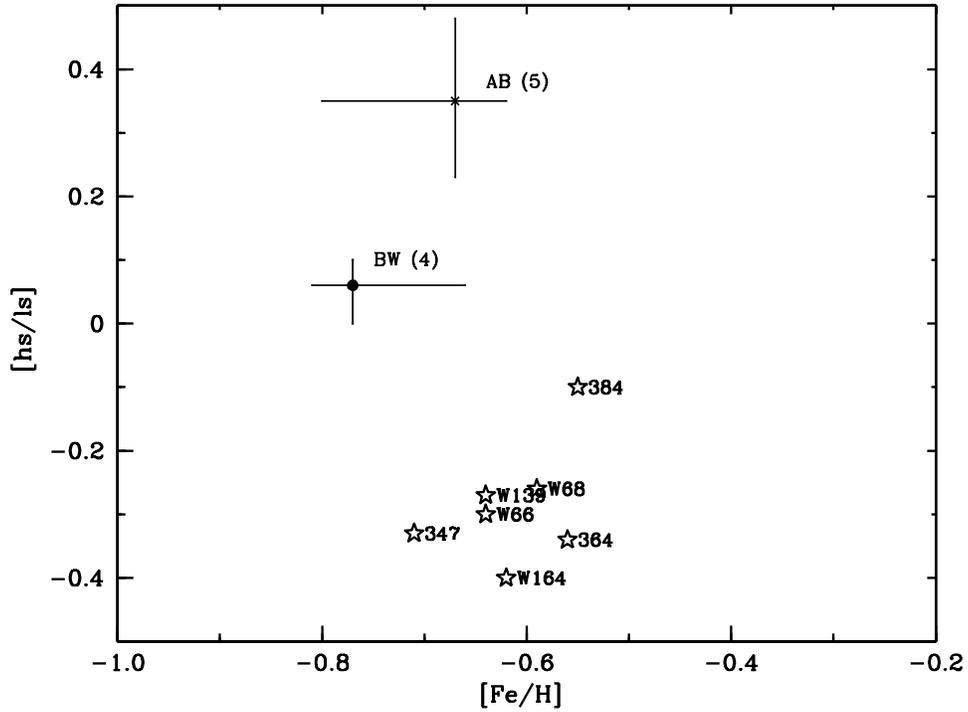}
\caption{Abundance ratio [hs/ls] as a function of metallicity.  The published range of the RGB stars is shown while the program stars are labeled.  Note that W66 and W68 are RGB stars, as defined by the color-magnitude diagram (Figure \ref{hrd}).  The program stars clearly lie below and to the right of the previously studied RGB stars.  However, a real spread in the [hs/ls] in the cluster as a whole may explain this positioning. \label{fehhsls}}
\end{figure}

\clearpage

\begin{figure}
\epsscale{0.8}
\plotone{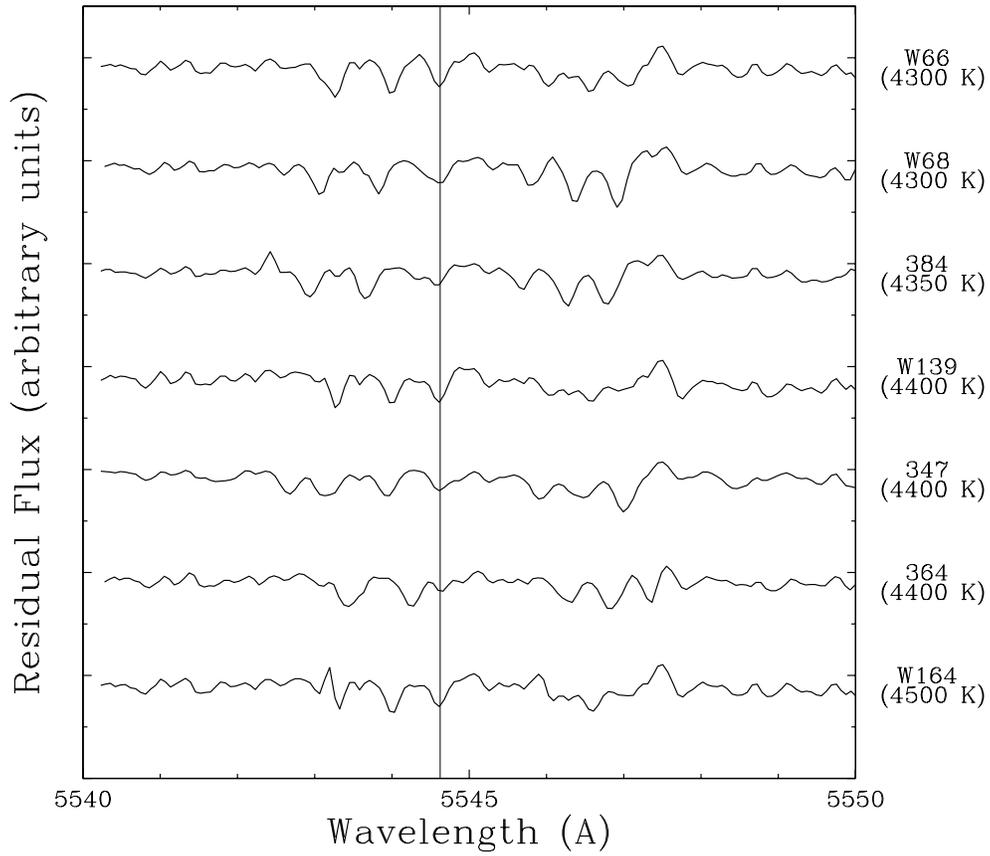}
\caption{The  Y I 5545\AA\ region shown in order of increasing effective temperature.  This plot suggests that there is no clear trend with effective temperature, suggesting that the observed variation is real.\label{ywitht}}
\end{figure}

\clearpage

\begin{figure}
\epsscale{0.8}
\plotone{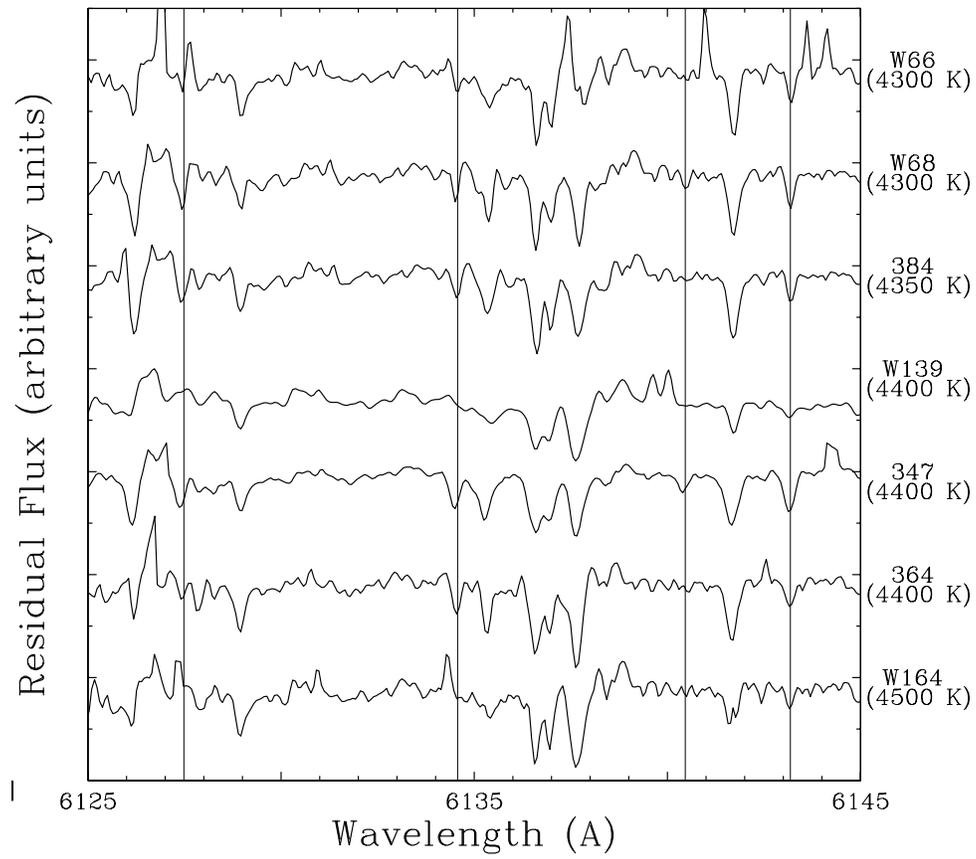}
\caption{As for Figure \ref{ywitht} except for the Zr I 6140\AA\ region.\label{zrwitht}}
\end{figure}

\clearpage

\begin{deluxetable}{cccc}
\tablewidth{0pt}
\tablecaption{Element abundances from previous studies of RGB stars in 47 Tuc.\label{wall47tuc}}
\tablehead{
\colhead{}&\colhead{Average BW92}&\colhead{Average AB05}&\colhead{Average J04}\\
\colhead{}&\colhead{4}&\colhead{5}&\colhead{12}}
\startdata
[Fe/H]  & -0.81 & -0.67 & -0.69\\ \hline
   & [X/Fe] & [X/Fe] & [X/Fe] \\
Na & +0.11 &+0.03&\nodata\\
Sr & \nodata & \nodata & +0.36 \\
Y & +0.45 & \nodata& -0.11	 \\ 
Zr & -0.39 &-0.17  & \nodata \\
Ba &\nodata& +0.31 & +0.35\\
La & +0.18 & +0.05 & \nodata\\
Eu & +0.25 &+0.33 & +0.17	 \\ \hline
hs/ls & +0.06  & +0.35 & +0.23  \\
\enddata
\end{deluxetable}

\clearpage 

\begin{deluxetable}{cccccccc}
\tablewidth{0pt}
\tablecaption{Observational details of program stars in 47 Tuc.\label{obsstars}}
\tablehead{
\colhead{Star}& \colhead{V}&\colhead{J}&\colhead{K}&\colhead{RA}&\colhead{Dec}&\colhead{Exposure Time} &\colhead{S:N}\\
\colhead{}&\colhead{(a)}&\colhead{(b)}&\colhead{(b)}&\colhead{ h m s}&\colhead{$^\circ$ m s}&\colhead{s}&\colhead{}}
\startdata
W66      &12.55	 &10.21   &9.31   & 00 22 13.8 &    -72 16 12 & 3 x 1800 & 45\\
W68      &12.30$^1$	 &10.17   &9.26   & 00 21 01.9 &    -72 16 15 &4 x 1800  &45\\					
W139     &12.70	 &10.33   &9.53   & 00 20 50.4 &    -72 21 10 & 5 x 1800& 40\\			
W164     &12.62	 &10.41   &9.57   & 00 20 40.8 &    -72 21 58 &4 x 1800 &40\\							
347      &12.19	 &9.66    &8.75   & 00 22 52.2 &    -72 26 13 &2 x 1800 &40\\
364      &12.45$^2$	 &10.23   &9.38   & 00 21 29.6 &    -72 28 13 &4 x 1800  &40\\						  
384      &12.29	 &9.98    &9.10   & 00 21 15.0 &    -72 22 59 & 3 x 1800&40\\		
\enddata
\tablenotetext{a}{\citet{1978AJ.....83..376C}}
\tablenotetext{b}{2MASS All-Sky Point Source Catalog at www.irsa.ipac.caltech.edu}
\tablenotetext{1}{\citet{1961ApJ...133..430W}}
\tablenotetext{2}{\citet{1977A&AS...27..381L}}
\end{deluxetable}

\clearpage

\begin{deluxetable}{lccccccccccc}
\rotate
\tablewidth{0pt}
\tablecaption{Atmospheric Parameters for Program Stars.\label{atmosparam}}
\tablehead{
\colhead{Star}  & \colhead{V} &\colhead{J}&\colhead{K} &\colhead{Teff (K)} & \colhead{Teff (K)}&\colhead{Teff (K)}& \colhead{log g} & \colhead{log g}&\colhead{$\nu$} & \colhead{CN} & \colhead{}\\
\colhead{}&\colhead{(a)} & \colhead{(b)}&\colhead{(b)}&\colhead{Spec}&\colhead{Phot(B-V)}&\colhead{Photo(J-K)}&\colhead{Spec}&\colhead{Photo}&\colhead {m $s^{-1}$}& \colhead{(c)}&\colhead{(d)}}
\startdata
W66      &12.55	 &10.21  &9.31      &4300 &4150 &3900 &1.0 &1.0 &1.9 & Weak &\nodata \\
W68 &12.30$^1$	 &10.17  &9.26      &4300 &4060 &3900 &1.3 &1.0 &1.3 & Strong &\nodata\\
W139     &12.70	 &10.33  & 9.53     &4400 &4300 &4130 &1.5 &1.1 &1.9 & Strong &\nodata\\
W164     &12.62	 &10.41  &9.57      &4500 &4450 &4040 &1.3 &1.1 &2.4 &  Absent &\nodata\\
347      &12.19	 &9.66   &8.75      &4400 &4210 &3900 &1.5 &1.0 &1.7 &  Absent & Strong\\
364  &12.45$^2$	 &10.23  &9.38      &4400 &4250 &4020 &1.3 &1.1 &2.0 &  Absent & \nodata\\
384      &12.29	 &9.98   &9.10      &4350 &4220 &3960 &1.5 &1.1 &1.7 &  \nodata & \nodata\\
\enddata
\tablenotetext{a}{\citet{1978AJ.....83..376C}}
\tablenotetext{b}{2MASS All-Sky Point Source Catalog at www.irsa.ipac.caltech.edu}
\tablenotetext{c}{\citet{1978A&A....70..115M}}
\tablenotetext{d}{\citet{1979ApJ...230L.179N}}
\tablenotetext{1}{\citet{1961ApJ...133..430W}}
\tablenotetext{2}{\citet{1977A&AS...27..381L}}
\end{deluxetable}

\clearpage

\begin{deluxetable}{lccccc}
\tabletypesize{\scriptsize}
\tablewidth{0pt}
\tablecaption{Heavy element lines used in order of increasing excitation potential.\label{linesused}}
\tablehead{
\colhead{Element} & \colhead{Wavelength} & \colhead{$\chi$} & \colhead{Wylie log gf} & \colhead{Brown log gf}&\colhead{Published Lab Values(a)}}
\startdata
Y I & 6435.05 &0.07& -1.02 & -0.82 &-0.82 \\
    & 5526.73&2.00&-0.65 & &\nodata \\
Y II & 5119.12 &0.99 & -1.36 &&-1.36 \\
     & 5087.43 &1.08 & -0.17 &&-0.17 \\  
     & 5473.39 &1.73 &-1.02 && \\
     & 5544.62 &1.74 & -1.19 &&-1.09 \\
     &5546.03  &1.75 & -1.10 &&-1.10 \\
Zr I&6134.57 &0.00 & -1.28 &-1.28 &-1.28 \\
    & 6143.18 &0.07 & -1.10 &-1.10& -1.10  \\
    &6127.48 &0.15 & -1.06 &-1.06& -1.06 \\
    & 6140.46 &0.52 & -1.41 &-1.41& -1.41  \\
    &4815.64 &0.60 & -0.13 & &-0.03  \\
    &4828.06 &0.62 & -0.64 & &-0.64 \\
    & 4815.06 &0.65 & -0.53 && -0.53  \\
    & 4805.89 &0.69 & -0.57 && -0.42  \\
    & 4809.48 &1.58 & +0.16 && +0.16  \\
Zr II & 5112.28 &1.66 & -0.59 & &-0.59  \\
      &5350.36 &1.77 & -1.18 & &\nodata  \\
      & 5350.09 &1.83 & -0.94 && \nodata  \\
La II   &5808.31&0.00 &-1.56 &-2.34 &\nodata \\
	&5805.77 &0.13 & -1.61 & &-1.56  \\
      &5114.56&0.24&-1.06 & &\nodata \\
      &5797.60 &0.24 & -1.51 && -1.03  \\
      & 5122.99 &0.321 &-0.93 & & \\
Nd II  & 5167.92 &0.56  &  -0.980 & &\nodata \\
 & 5165.13 & 0.68 &   -0.060 & &\nodata \\
 & 5161.71 &0.74  &   -0.980 & &\nodata \\
 &5804.00 &0.74 & -0.53 & &-0.53  \\
      &5795.15 &1.26 & -1.13 & &-1.13  \\
Eu II &6645.13 &1.38 & +0.40 &+0.20 &\nodata  \\
\enddata
\tablenotetext{a}{Y: \citet{1981ApJ...248..867B}, Zr: \citet{2001ApJ...556..452L}, La: \citet{2003ApJS..148..543D} and Nd: \citet{1982ApJ...261..736H}}
\end{deluxetable}

\clearpage

\begin{deluxetable}{c|ccccccccccccccc}
\rotate
\tabletypesize{\scriptsize}
\tablewidth{0pt}
\tablecaption{47 Tucanae Abundances.\label{47tucres}}
\tablehead{
\colhead{} & \colhead{}  & \colhead{W66} &  \colhead{}&\colhead{W68} &  \colhead{}& \colhead{W139}  & \colhead{} &  \colhead{W164}& \colhead{}& \colhead{347} &\colhead{} &\colhead{364} &\colhead{} &\colhead{384}&\colhead{}\\
\colhead{Species}&\colhead{log$\epsilon_\odot$(X)}&\colhead{N}&\colhead{[Fe/H]}&\colhead{N}&\colhead{[Fe/H]}&\colhead{N}&\colhead{[Fe/H]}&\colhead{N}&\colhead{[Fe/H]}&\colhead{N}&\colhead{[Fe/H]}&\colhead{N}&\colhead{[Fe/H]}&\colhead{N}&\colhead{[Fe/H]}}
\startdata
FeI & 7.52 &20 &-0.63 &21 &-0.60 &24 &-0.62 &25 &-0.60 &25 &-0.66 &28 &-0.56 &23 &-0.55 \\
FeII & 7.52 &8 &-0.65 &8 &-0.57 &8 &-0.69 &9 &-0.63&6 &-0.76 &8 &-0.56 &8 &-0.55  \\ \hline
    &      &   & [X/Fe]& & [X/Fe]& & [X/Fe]& & [X/Fe]& & [X/Fe]& & [X/Fe]& & [X/Fe]\\
NaI & 6.33 &2  & 0.77 &2  & 1.04 &2  & 0.31 &2  &0.51  &2  &0.60  &2  &0.55  &2  &0.74 \\
YI & 2.24 &2 &0.66 &2 &0.68 &2 &0.80 &2 &0.73&1 &0.89 &1 &0.49 &2 &0.28  \\
YII & 2.24 &4 &0.68 &4 &0.60 &5 &0.72 &5 &0.76 &4 &0.89 &4 &0.61 &5 &0.28 \\
ZrI & 2.60 &6 &0.58 &6 &0.73 &5 &0.60 &6 &0.63 &7 &0.93 &7 &0.75 &4 &0.58 \\
ZrII & 2.60 &2 &0.58 &2 &0.55 &2 &0.82 &2 &0.73  &2 &0.64 &2 &0.57 &2 &0.28\\
LaII & 1.22 &4 &0.28 &5 &0.27 &3 &0.44 &4 &0.21  &3 &0.46 &5 &0.21 &5 &0.29 \\
NdII & 1.50 &3 &0.38 &4 &0.50 &4 &0.50 &4 &0.44 &3 &0.56 &4 & 0.32 &4 &0.23 \\
EuII & 0.51 &1 &0.22 &1 &0.04 &1 &0.26 &1 &0.20&1 &0.23 &1 &-0.08 &1&0.12\\\hline
ls/Fe & & & 0.63 & & 0.64 & & 0.74 & & 0.72 & & 0.84 & & 0.61 & & 0.36 \\
hs/Fe & & & 0.33 & & 0.39 & & 0.47 & & 0.33 & & 0.51 & & 0.27 & & 0.26 \\
hs/ls&     &  &-0.30   & &-0.26 & &-0.27 & &-0.40 & &-0.33 & &-0.34 & &-0.09 \\
\enddata
\end{deluxetable}

\clearpage

\begin{deluxetable}{ccccc}
\tablewidth{0pt}
\tablecaption{Mean abundances and uncertainties for the whole sample, and the sample of confirmed AGB stars.\label{sigmas}}
\tablehead{
\colhead{Species}&\colhead{Average}&\colhead{$\sigma$}&\colhead {Average }&\colhead{$\sigma$}\\
\colhead {}      & \colhead{7 stars} &\colhead{}&\colhead{5 stars}&\colhead{}\\
\colhead {}      & \colhead{[Fe/H]}  & \colhead {}&\colhead {[Fe/H]}&\colhead {}}
\startdata
Fe I & -0.61 &$\pm$0.06& -0.57 &  $\pm$0.06\\
Fe II & -0.64 &$\pm$0.10 & -0.58 & $\pm$0.10\\\hline
     & [X/Fe] & &[X/Fe] & \\
Na I & +0.67 & $\pm$0.23 & +0.54 & $\pm$0.10\\
Y I &+0.63 &$\pm$0.20 & +0.60 & $\pm$0.20\\
Y II & +0.65 &$\pm$0.18 & +0.57 & $\pm$0.18\\
Zr I & +0.69 &$\pm$0.15 & +0.60 & $\pm$0.14\\
Zr II & +0.61 &$\pm$0.16  & +0.56 & $\pm$0.16 \\
La II & +0.31 &$\pm$0.19 & +0.26 & $\pm$0.19\\
Nd II & +0.42 &$\pm$0.13 & +0.38 & $\pm$0.12\\
Eu II & +0.15 &$\pm$0.12 & +0.13 & $\pm$0.10\\ \hline
ls/Fe & 0.65 &$\pm$0.10 & +0.59 & $\pm$0.10\\
hs/Fe & 0.37 & $\pm$0.14 & +0.32 & $\pm$0.14\\
hs/ls &-0.28 &$\pm$0.11 & -0.27 & $\pm$ 0.11\\
\enddata
\end{deluxetable}

\begin{deluxetable}{cccc}
\tablewidth{0pt}
\tablecaption{Abundance dependence on atmospheric parameters.\label{deponparam}}
\tablehead{
\colhead{Species}&\colhead{$\Delta$T$_{eff}$}&\colhead{$\Delta$log g}&\colhead {$\Delta \xi$}\\
\colhead {}      & \colhead{+100 K} &\colhead{+0.5}&\colhead{+0.5 km s$^{-1}$}}
\startdata
Fe I & +0.06 & +0.15 & -0.16 \\
Fe II & -0.12 & +0.36 & -0.18 \\
Y I & +0.21 & +0.03 & -0.05 \\
Y II & -0.01 & +0.27 & -0.12 \\
Zr I & +0.28 & +0.09 & -0.07 \\
Zr II & -0.01 & +0.25 & -0.07 \\
La II & +0.05 & +0.25 & -0.05 \\
Nd II & -0.04 & +0.24 & -0.05 \\
Eu II & 0.00 & +0.26 & -0.07 \\ 
\enddata
\end{deluxetable}

\end{document}